\documentstyle[prl,aps,multicol,epsf]{revtex}
\large
\begin{document}
\draft
\title{Josephson Frequency Singularity in the Noise of \\
Normal Metal -- Superconductor Junctions}
\author{Gordey~B.\ Lesovik$^{a,b}$, 
Thierry Martin $^{a}$ and Julien Torr\`es $^{a}$}
\address{$^{a}$Centre de Physique Th\'eorique,
Universit\'e de la M\'editerran\'ee,
Case 907, F-13288 Marseille Cedex 9, France}
\address{$^{b}$Institute of Solid State Physics, Chernogolovka 142432,
 Moscow district, Russia}
\maketitle
\begin{abstract}
A singularity at the Josephson frequency
in the noise spectral density of a disordered normal metal -- 
superconductor junction is predicted for bias voltages
below the superconducting gap. 
The non-stationary Aharonov-Bohm effect, recently
introduced for normal metals  
\cite{Lesovik Levitov}, is proposed as a tool for  
detecting this singularity. In the presence 
of a harmonic external field, the derivative of the noise 
with respect to the voltage bias reveals jumps when the 
applied frequency is commensurate 
with the Josephson frequency associated with
this bias. The height of these jumps is 
non-monotonic in the amplitude of the periodic field.
The superconducting flux quantum enters this dependence.
%%%%%%%%%%%%%%%%%%%%%%%%%%%%%%%%%%%%%%%%%%%%%%%%%%%%%%%%
Additional singularities in the frequency dependent noise
are predicted above gap.
\end{abstract}
\begin{multicols}{2}
\narrowtext


The observation of the Josephson effect \cite{Josephson}
typically requires two superconductors in contact.  
Here, we present two situations where a Josephson 
frequency can be observed in a normal metal -- superconductor 
(NS) junction. Although the Josephson frequency does not 
manifest itself in the average current through 
the NS junction, the 
noise characteristics bears clearly its signature. 
Superconducting features have been predicted in NS junctions: 
the doubling of the shot noise \cite{Khlus}
and the crossover from thermal noise to excess noise
at $2k_BT= (2e)V$ ($V$ applied bias) are examples 
\cite{Khlus,Martin}. Recent experiments \cite{Sanquer}
in SNS junctions seem to be in qualitative agreement 
with this crossover.
First we present a general framework to investigate
the finite frequency noise of NS junctions, 
second, we propose an equivalent zero frequency, 
constant bias measurement in the presence of an 
harmonic perturbation to detect this analog of the 
Josephson effect. 
%we go further in the investigation of NS transport 
%by calculating finite frequency noise at constant bias
%and zero frequency noise in the presence of an harmonic 
%perturbation. 
In the first case, the singularity at the 
Josephson frequency originates from the 
time oscillations of the current--current correlation 
function. In the second scheme, the superposition of an
alternating field to the bias voltage leads to 
steps/singularities in the noise derivatives as 
a function of DC voltage. The strength of these singularities
is non-monotonic with the amplitude of the harmonic 
perturbation. This so called
``non-stationary Aharonov--Bohm effect'' has been predicted 
\cite{Lesovik Levitov} and observed experimentally 
\cite{Schoellkopf 2} in normal mesoscopic samples.
Similarly, the singular behavior of finite frequency
noise is understood both theoretically and 
experimentally \cite{Yang,Schoellkopf frequency}.   
For an NS junction with a bias smaller than the gap, 
the physical quantities can in principle be obtained 
from the normal metal results by replacing everywhere 
the electron charge by the charge of a Cooper pair. 
In both the conductance
and the excess noise of NS junctions, this doubling of 
the electron charge has to be divided by two in the 
prefactors, in order to account 
for the lack of spin degeneracy. However, no 
detailed calculation of these effects for finite 
frequencies is available, and the generalization to 
above gap voltages cannot be obtained from
an effective carrier charge.   

\begin{figure}
\epsfxsize 8.5 cm
\centerline{\epsffile{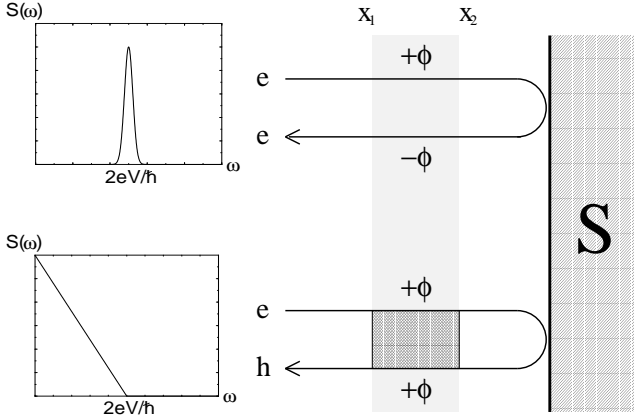}}
\medskip
\caption{\label{Fig1}
a) Zero order spectral density of noise
for the non-stationary Josephson effect (top) with a finite linewidth,
and for the NS junction (bottom). b) The NS boundary. The light
shaded region determines where the electron/hole wave function
may accumulate phase: schematic description of the two 
scattering processes, Andreev and normal reflection.}
\end{figure}

A general framework is provided for both 
the stationary and non-stationary problems. 
Consider the coherent normal metal -- superconductor 
junction in Fig. \ref{Fig1}. A steplike dependence of the gap is 
assumed at the boundary, and the one channel states for 
electrons and holes incident from the normal side
are specified by a scattering matrix with elements
$s_{\alpha\beta}$ ($\alpha,\beta=e,h$). 
A time dependent phase $\Phi(t)$ is accumulated on 
the normal side by electrons impinging on the 
superconductor (incident holes will bear the opposite 
phase). Normally reflected particles will not accumulate 
this external phase. No inelastic effects other than this perturbation
are assumed close to the NS boundary.
The probability for 
Andreev reflection \cite{Andreev} is $R_A=|s_{he}|^2$.
In general, a time dependent 
formulation \cite{Bogolubov time} of the Bogolubov--de Gennes (BdG) 
equations is needed. Here, we assume that the
time dependence is adiabatic, so that appropriate
elements of the scattering matrix are simply multiplied 
by $\exp[\pm2i\Phi(t)]$ in the BdG equations \cite{BdG}. 
The electron and hole wave functions are given by: 
\begin{eqnarray}
\nonumber
u_{\alpha} (x,t)&\simeq & \Bigl[
  \delta_{\alpha e} \left( e^{ik_+x} + s_{ee}
e^{-ik_+x} \right) 
\\ && \hspace{5mm}
+ \delta_{\alpha h} s_{eh} e^{-2i\Phi(t)} e^{-ik_+x}
\Bigr]/\sqrt{h v_+}~,
\label{u BdG}
\\
\nonumber
v_{\alpha} (x,t)&\simeq& \Bigl[
  \delta_{\alpha h} \left( e^{-ik_-x} + s_{hh} 
e^{ik_-x} \right)
\\ && \hspace{5mm}
+ \delta_{\alpha e} s_{he} e^{ 2i\Phi(t)} e^{ik_-x}
\Bigr]/\sqrt{h v_-}~,
\label{v BdG}
\end{eqnarray}
where $v_\pm=\hbar k_\pm/m$ is the velocity of waves
with $\hbar k_\pm=\sqrt{2m(\mu_S\pm \varepsilon)}$. 
In what follows, we neglect the difference between 
the hole and electron wave numbers in Eq. 
(\ref{u BdG}) and (\ref{v BdG}), assuming the 
chemical potential $\mu_S$ to be large. 

Performing the Bogolubov
transformation on the current operator, the statistical 
average of the current--current correlator is computed: 
%\cite{Lesovik Golubov}:
\begin{eqnarray}
\nonumber
&&
\langle \langle I(t_1)I(t_2) \rangle \rangle =
\frac{e^2 \hbar^2}{2m^2 }
\sum_{\alpha,\beta} \int_0^{+\infty} \int_0^{+\infty} d\varepsilon d\varepsilon' \Big\{
\hspace{1.8cm}
\\
\nonumber
&& \hspace{0.5cm}
  f_{\alpha}(1-f_{\beta})     e^{i (\varepsilon'- \varepsilon)(t_2-t_1)/\hbar}
\\
\nonumber
&& \hspace{1cm}
\Bigl[
  (u_{\beta} \!\! \stackrel{\leftrightarrow}{\partial} \!\! u_{\alpha}^*)_{t_1} (u_{\beta}^* \!\! \stackrel{\leftrightarrow}{\partial} \!\! u_{\alpha})_{t_2} 
+ (v_{\beta} \!\! \stackrel{\leftrightarrow}{\partial} \!\! v_{\alpha}^*)_{t_1} (v_{\beta}^* \!\! \stackrel{\leftrightarrow}{\partial} \!\! v_{\alpha})_{t_2} 
\\
\nonumber
&& \hspace{1.5cm}
+ (u_{\beta} \!\! \stackrel{\leftrightarrow}{\partial} \!\! u_{\alpha}^*)_{t_1} (v_{\beta}^* \!\! \stackrel{\leftrightarrow}{\partial} \!\! v_{\alpha})_{t_2} 
+ (v_{\beta} \!\! \stackrel{\leftrightarrow}{\partial} \!\! v_{\alpha}^*)_{t_1} (u_{\beta}^* \!\! \stackrel{\leftrightarrow}{\partial} \!\! u_{\alpha})_{t_2} 
\Bigr]
\\
\nonumber
&& \hspace{0.5cm}
+ f_{\alpha} f_{\beta}        e^{-i (\varepsilon + \varepsilon')(t_2-t_1)/\hbar}
\\
\nonumber
&& \hspace{1cm}
(u_{\beta}^* \!\! \stackrel{\leftrightarrow}{\partial} \!\! v_{\alpha}^*)_{t_1} \Bigl[
(u_{\alpha} \!\! \stackrel{\leftrightarrow}{\partial} \!\! v_{\beta})_{t_2} + (u_{\beta} \!\! \stackrel{\leftrightarrow}{\partial} \!\! v_{\alpha})_{t_2}
\Bigr]
\\
\nonumber
&& \hspace{0.5cm}
+ (1-f_{\alpha})(1-f_{\beta}) e^{i (\varepsilon + \varepsilon')(t_2-t_1)/\hbar}
\\
&& \hspace{1cm}
\Bigl[
(u_{\alpha} \!\! \stackrel{\leftrightarrow}{\partial} \!\! v_{\beta})_{t_1} + (u_{\beta} \!\! \stackrel{\leftrightarrow}{\partial} \!\! v_{\alpha})_{t_1}
\Bigr]
(u_{\beta}^* \!\! \stackrel{\leftrightarrow}{\partial} \!\! v_{\alpha}^*)_{t_2} 
\Bigr\}
~,
\label{current current}
\end{eqnarray}
where
$u \!\! \stackrel{\leftrightarrow}{\partial} \!\! v
= u \partial_x v - v \partial_x u$,
and $\langle\langle \hspace{2mm} \rangle\rangle$ means that the 
square of the average current has been subtracted.
The time dependence indicated with the indices $t_1$ and $t_2$
is explicited in the electron and hole wave functions of 
Eq. (\ref{u BdG}) and (\ref{v BdG}).
$f_e(\varepsilon)=f(\varepsilon-eV)$ 
($f_h(\varepsilon)=1-f(-\varepsilon-eV)$) denote the electron (hole) 
distribution function and $f$ is the Fermi--Dirac function.
While only the terms containing the product 
$f_\alpha(1-f_\beta)$ contribute to the low frequency 
noise \cite{Beenakker,Datta}, terms proportional
to $f_\alpha f_\beta$ and $(1-f_\alpha)(1-f_\beta)$,
which involve matrix elements between states which differ
by two quasiparticles,
contribute to the finite frequency noise and to the 
non-stationary calculation below. 
The current--current correlator of Eq. (\ref{current current})
would have a spatial dependence if the exact wave vector 
of electrons and holes was included in this finite frequency 
calculation \cite{Buttiker frequency}. However, this spatial 
dependence is only relevant at frequencies large 
compared to the inverse time of flight through 
the sample, which are not considered here

In order to describe non-stationary situations,
we introduce the double Fourier transform  of the   
current--current correlator:
\begin{equation}
\widetilde{S}(\Omega_1,\Omega_2) =
\int \int dt_1 dt_2 e^{i \left( \Omega_1 t_1 + \Omega_2 t_2 \right)}
\langle \langle I(t_1)I(t_2) \rangle \rangle
~.
\label{general Fourier transform}
\end{equation}

If the translational invariance in time is broken either 
by an external alternating field (with frequency $\Omega$) 
or spontaneously, like in the non-stationary 
Josephson effect (NJE) \cite{Josephson},
this double Fourier transform can be written as: 
\begin{equation}
\widetilde{S}(\Omega_1,\Omega_2) =
\sum_{m=-\infty}^{+\infty}  
2\pi \delta(\Omega_1+\Omega_2-m\Omega) S^{(m)} (\Omega_2)
~.
\label{NSJ}
\end{equation}

In both cases the zero harmonic, which is proportional
to $\delta (\Omega_1 +\Omega_2)$
is the most standard quantity to study.
In the presence of this time invariance this is 
the only term which is present.
For the above mentioned NJE, the zeroth order
spectral density $S^{(0)}(\omega ) $, defined in Eq. (\ref{NSJ})
is proportional to the spectral function line-shape $D(\omega)$,
shifted by the Josephson frequency: 
$S^{(0)}(\omega )=D(\omega - 2eV/\hbar )$
which is depicted in Fig. 1 a).  

We first consider the finite frequency noise in the 
presence of a constant bias. Because of translational 
invariance in time, the Fourier transform reduces to:
\begin{equation}
\nonumber
\widetilde{S}(\Omega_1,\Omega_2)=2 \pi \delta(\Omega_1 + \Omega_2)S(\Omega_2)
~.
\end{equation}
where $S(\Omega_2)$ is the ``usual'' frequency dependent noise. 
In what follows, we assume that the energy dependence of the 
scattering matrix coefficients can be neglected for biases 
small compared to the gap. At arbitrary 
temperature and frequency below the superconducting gap, the noise is:
\begin{eqnarray}
\nonumber
S(\omega) \! &=& \!
\frac{8e^2}{h} R_A^2 \! \! \int_{-\infty}^{+\infty} \! \! d\varepsilon f(\varepsilon-eV) \! \left( 1 - f(\varepsilon - eV - \hbar \omega) \right)
\\
&& \hspace{5mm}
+ \frac{4e^2}{h} R_A (1- R_A) F_V(\omega)\label{noise frequency temp}
~,
\end{eqnarray}
with
\begin{eqnarray}
\nonumber
F_V(\hbar\omega) &=&
 \int_{-\infty}^{+\infty} d\varepsilon \, \Bigl[
  f(\varepsilon-eV) \left( 1 - f(\varepsilon + eV - \hbar \omega) \right)
\\
&& \hspace{0.8cm}
+ f(\varepsilon + eV +\hbar\omega) \left( 1 - f(\varepsilon - eV) \right)
\Bigr]
~.
\end{eqnarray}
Note that the (constant) phase factor $\Phi$ does not appear in the 
above because the geometry is open.
The first term in Eq. (\ref{noise frequency temp}) is important
when thermal fluctuations are present. At zero temperature and for positive
frequencies, only the second term contributes, this gives: 
\begin{equation}
S(\omega) = \frac{4e^2}{h} R_A \left( 1 - R_A \right) (2eV - \hbar \omega) \; \theta (2eV - \hbar \omega)
~.
\label{zero temperature noise}
\end{equation}
with the convention $eV>0$.
The striking feature in Eq. (\ref{zero temperature noise}) is the 
singularity at the Josephson frequency $2eV/\hbar$. This frequency 
scale appears in this normal superconducting geometry because 
below the gap, the only available charge transfer process
involves the conversion of electrons into holes. To gain a better 
understanding of this singularity, remember that 
for a junction between two superconductors, the order 
parameter on each side oscillates as $\exp[-i2\mu_{S_{1,2}}t/\hbar]$
with $\mu_{S_1}$ and $\mu_{S_2}$ the chemical potentials on each side, 
leading to a current oscillation with frequency $2(\mu_{S_2}-\mu_{S_1})/\hbar$.
For a junction between two normal metals, where the electrons
wave functions on each side oscillate like $\exp[-i\mu_{1,2} t/\hbar]$,
a singularity in noise at 
$(\mu_{2}-\mu_{1})/\hbar$ has been pointed out 
\cite{Yang}. In the case of an NS junction, 
the electron and hole 
wave functions on the normal side have a time dependence 
$\exp[-i(\mu_S\pm eV)t/\hbar]$ (where $\mu_S$ is the 
chemical potential of the superconductor), leading to the 
singular behavior in Eq. (\ref{zero temperature noise}).
A similar but weaker singularity has been pointed out in the 
fractional quantum Hall regime (FQHE) \cite{Wen}, at the 
``Josephson'' frequency $e^*V/\hbar$ with $e^*/e$ the electron
filling factor.   

At $\omega=0$, we recover the doubled shot noise of NS junctions 
although the current--current correlation 
function in Eq. (\ref{general Fourier transform}) 
has not been symmetrized here. At zero frequency, the 
choice of considering $\langle \langle I(t_1)I(t_2) \rangle \rangle$ 
or its symmetrized 
analog does not matter. At finite frequency,
depending on the measurement procedure, both correlators may 
occur \cite{Loosen}. Here, the time dependence 
of the symmetrized correlator at $t>\hbar/eV$ is:
\begin{equation}
\langle \langle I(t)I(0)+I(0)I(t) \rangle \rangle \! =
\frac{8e^2}{\pi^2}R_A \left( 1 - R_A \right){\sin^2(eVt/\hbar)\over t^2}
~.
\label{explicit time}
\end{equation}
Up to our knowledge experimental techniques do not allow for the direct 
observations of these oscillations in time. Nevertheless,
finite frequency measurements are possible 
\cite{Schoellkopf frequency}.

We now turn to a situation where a time dependent vector potential 
is applied near the boundary (Fig. 1). 
Because low frequency measurements are more accessible than 
finite frequency ones, we suggest that a
zero frequency noise analysis of a sinusoidally perturbed 
system is a more straightforward tool to 
investigate the presence 
of the Josephson frequency in this NS system.
In addition, this setup gives us an extra opportunity to
observe a superconducting behavior, i.e. a doubling 
of the electron charge, as shown below.   
The phase is chosen to be
a periodic function of time $\Phi(t)=\Phi_a\sin(\Omega t)$ with
$\Phi_a\equiv 2\pi\int_{x_1}^{x_2} dx A_x/\phi_0$ where $\phi_0=hc/e$ 
is the normal flux quantum, $[x_1,x_2]$ defines the interval
where the vector potential is confined (Fig. \ref{Fig1}). 
The effect of this perturbation on the average 
current is straightforward in the limit where $R_A$ depends 
weakly on the energy: it brings a periodic modulation of 
the current 
$\Delta I=(4e^2/h)R_A[\hbar\Omega/e]\Phi_a\cos(\Omega t)$.
In the current-current correlations, this modulation leads to a 
non-monotonic effect as a function of phase, in contrast with the 
electromotive force action on the current. Note that no closed 
topology is imposed, in contrast to the usual AB effect.
Nevertheless, the phase is periodically modulated, and 
yields a non-zero contribution in the noise 
harmonics despite the open geometry.

The low frequency noise is computed by first performing 
the time integrals in Eq. (\ref{general Fourier transform})
from the explicit time dependence of the current 
matrix elements. Using the generating function of the Bessel
functions $J_n$, one obtains:
\begin{eqnarray}
\nonumber
S^{(0)}(0) &=&
\frac{4e^2}{h} R_A (1- R_A) \sum_{m=-\infty}^{+\infty} J_m^2(2\Phi_a) F_V( m \hbar \Omega)
\\
&&
+\frac{8e^2}{h} R_A^2 k_B T 
~.
\label{Noise AB}
\end{eqnarray}
The phase $2\Phi$ (Fig. \ref{Fig1}) accumulated in the 
Andreev process leads to a factor $2$ in the argument 
of $J_n$, which is reminiscent of the Cooper pair charge. 
The temperature dependence in Eq. (\ref{Noise AB}) 
is specified by $F_V(m \hbar \Omega)= 
(2eV - m \hbar \Omega)\coth[(2eV-m \hbar \Omega)/2k_BT]$
determines how the steps in the noise
derivative
\begin{equation}
\frac{\partial S^{(0)}(0)}{\partial V} \simeq
\frac{8e^3}{h} R_A (1- R_A) \sum_{m=-M}^{+M} J_m^2(2\Phi_a)
\label{derivative}
\end{equation}
are smeared with temperature. In Eq. (\ref{derivative}) 
the sum over harmonics has a cutoff at 
$M = \lfloor 2eV/\hbar \Omega \rfloor$. 
In experiments \cite{Schoellkopf 2} it is more convenient to characterize the 
non-monotonic dependence on voltage by taking the second 
derivative of the AB contribution to the noise. This is illustrated
for two distinct temperatures in Fig. 2. 
\begin{figure}
\epsfxsize 8.5 cm
\centerline{\epsffile{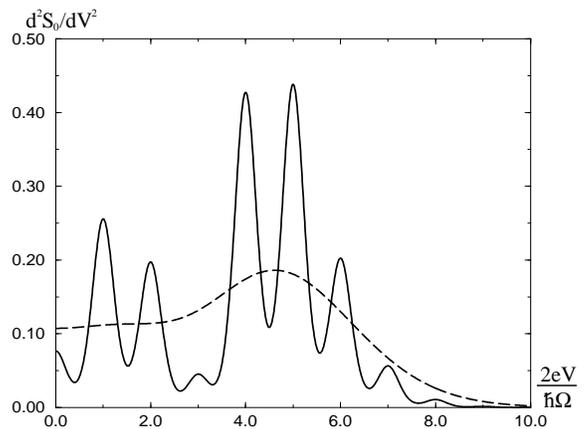}}
\medskip
\caption{\label{Fig2}
Non-stationary AB effect: plot of 
$\partial^2S/\partial V^2$, expressed in units of 
$(8e^4/\pi\hbar^2\Omega)R_A(1-R_A)$, as a function of $2eV/\hbar\Omega$,
with the choice $\Phi_a=3$.
For $2k_B T=0.2\hbar\Omega$ (full line) and for 
$2k_B T=\hbar\Omega$ (dashed line)}
\end{figure} 
For small temperatures
$\hbar\Omega>2k_B T$, one observes oscillations as a function of
$2eV/\hbar\Omega$, with a clustering of large amplitude
peaks. 
For temperatures $\hbar\Omega<2k_B T$, although individual 
peaks can no longer be identified, clusters of 
of ``large'' steps in the noise derivative
continue to give an average contribution  to the non-stationary 
AB effect. This robustness enhances the likelihood
of experimental observation, which is addressed below. 

In Ref.  \cite{Schoellkopf 2},
a small alternative microwave voltage $V_1(t)=V_1\sin ( \Omega t)$ 
was superposed to the DC bias instead 
of applying locally a magnetic flux. If the drop in 
voltage occurs at the NS boundary then gauge invariance states 
that this is equivalent to the present proposal, 
with the substitution $\Phi_a=eV_1/\hbar\Omega$. 
If this drop is more extended, the analogy is
restricted as the time of flight of electrons and holes 
through the relevant region is increased. 
Moreover, the experiments of Ref. \cite{Schoellkopf 2}
were performed in the diffusive regime, with a diffusion 
time $\tau_D$ of the order of the $\Omega^{-1}$. Although this seems 
to be at the border of applicability of the present approach,
the assumptions made in the superconducting case are similar,
so one expects these structures to be rather robust. 
Note that for these predictions to be valid, the harmonic 
perturbation should not affect significantly the distribution 
function of electrons in the reservoirs.   
In addition, for $\Delta\gg eV$, a multichannel generalization of these 
results is obtained straightforwardly with the 
substitution $ R_A(1-R_A)= \sum_n R_{A_n}(1-R_{A_n})$, where 
$R_{A_n}$ are the eigenvalues of the Andreev reflection matrix
\cite{Martin,Martin Landauer,Beenakker,Datta}.

The present calculations are limited to the subgap 
regime, but can be readily extended to $eV>\Delta$
provided that one takes into account quasiparticle
scattering states in the superconductor. In turn, 
a specific model for the NS boundary has to be chosen
in order to specify the energy dependence of 
the scattering matrix coefficients.
Calculations using the BTK \cite{BTK}
model of a step pair potential
with a delta function impurity potential 
will be discussed elsewhere in detail \cite{Torres}. 
Results indicate additional singularities to the one at 
$\omega=2eV/\hbar$ due to the
presence of new scattering channels:
electron--like 
quasiparticle transmission and 
hole--like (Andreev) quasiparticle transmission in the 
superconductor, which give rise to cusps at 
$\omega=(eV\mp \Delta)/\hbar$. 
These occur because of the presence
of a sharp edge in the density of states of the 
superconductor. Note that contrarily to the subgap regime,
these striking features cannot 
be interpreted by the substitution of an 
effective charge. The above gap regime 
provides an additional justification for the 
systematic study presented here:
these frequency scales are tied to the energy intervals
which are relevant for a given bias voltage, rather than 
an effective charge.
Further increasing the bias $\Delta\ll eV$, the superconducting 
singularity $\omega=2eV/\hbar$ weakens, and the dominant 
singularities $(eV\pm\Delta)/\hbar\approx eV/\hbar$
give the analog of the singularity of 
normal metals. 

The non-stationary Aharonov--Bohm effect  
provides an extra tool for the study of dynamical effects 
in low frequency noise. Because the effective charge $2e$ appears both
the in the position of the steps in the noise derivative 
and in the magnitude of the steps, this method of analysis 
could be applied to other quantum fluids with an effective
charge which is different than $e$ because of 
collective excitations. Shot noise measurements
recently reported a direct measurement of the fractional
charge \cite{Saminadayar Reznikov}.  
The non-stationary 
Aharonov--Bohm effect could therefore also be 
applied for the investigation of fractional charge 
Josephson frequency signatures in the Hall effect.

Part of G.B.L.'s work was done in Zurich, 
supported by the Swiss NSF.   

\end{multicols}
\end{document}